\documentclass[journal]{IEEEtran}

\usepackage[utf8]{inputenc}
\usepackage[T1]{fontenc}
\usepackage{xspace}
\usepackage{url}
\usepackage{hyperref}
\usepackage[noabbrev, capitalize]{cleveref}
\usepackage{balance}

\usepackage{pifont}
\newcommand{\cmark}{\ding{51}}
\newcommand{\xmark}{\ding{55}}

\usepackage{amsmath, amssymb, dsfont}

\usepackage{graphicx}

\usepackage{diagbox} 
\usepackage{colortbl} 
\usepackage{multicol}  
\usepackage{multirow}
\usepackage{longtable, tabulary}
\usepackage{array}

\newcommand{\prometheus}{\textsc{PROMETHEUS}\xspace}
\newcommand{\prometheuslemma}{\textsc{PRO}cedural \textsc{MET}hodology for developing \textsc{HE}uristics of \textsc{US}ability\xspace}
\newcommand{\rrrc}{\textsc{R3C}\xspace}
\newcommand{\tasaUnicidad}{\ensuremath{\mathit{\Phi_P}}\xspace}

\newcommand{\setProblemas}[0]{\ensuremath{\mathds{P}}\xspace}
\newcommand{\problemasComunes}{\ensuremath{\mathit{P^*}}\xspace}
\newcommand{\problemasDominioUniq}{\ensuremath{\mathit{P_D}}\xspace}
\newcommand{\problemasControlUniq}{\ensuremath{\mathit{P_C}}\xspace}
\newcommand{\problemasDominio}{\ensuremath{\mathit{P^*_D}}\xspace}
\newcommand{\problemasControl}{\ensuremath{\mathit{P^*_C}}\xspace}
\newcommand{\tasaUnicidadAprox}{\ensuremath{\mathit{\Phi_P^{*}}}\xspace}

\newcommand{\dispersionDominioUniq}{\ensuremath{\mathit{\delta_D}}\xspace}
\newcommand{\dispersionControlUniq}{\ensuremath{\mathit{\delta_C}}\xspace}
\newcommand{\tasaDispersion}{\ensuremath{\mathit{\delta_P}}\xspace}

\newcommand{\severidadDominio}{\ensuremath{\mathit{\lambda_D^*}}\xspace}
\newcommand{\severidadControl}{\ensuremath{\mathit{\lambda_C^*}}\xspace}
\newcommand{\severidadDominioUniq}{\ensuremath{\mathit{\lambda_D}}\xspace}
\newcommand{\severidadControlUniq}{\ensuremath{\mathit{\lambda_C}}\xspace}
\newcommand{\tasaSeveridad}{\ensuremath{\mathit{\lambda_P}}\xspace}
\newcommand{\tasaSeveridadAprox}{\ensuremath{\mathit{\lambda_P^*}}\xspace}

\newcommand{\especificidadDominioUniq}{\ensuremath{\mathit{\varepsilon_D}}\xspace}
\newcommand{\especificidadControlUniq}{\ensuremath{\mathit{\varepsilon_C}}\xspace}
\newcommand{\tasaEspecificidad}{\ensuremath{\mathit{\varepsilon_P}}\xspace}

\newcommand{\heuristicasDominio}{\ensuremath{\mathit{H_D}}\xspace}
\newcommand{\heuristicasControl}{\ensuremath{\mathit{H_C}}\xspace}
\newcommand{\uc}{\ensuremath{\mathit{UC}}\xspace}
\newcommand{\ld}{\ensuremath{\mathit{LD}}\xspace}
\newcommand{\pd}{\ensuremath{\mathit{PD}}\xspace}
\newcommand{\up}{\ensuremath{\mathit{UP}}\xspace}
\newcommand{\dom}{\ensuremath{\mathit{D}}\xspace}
\newcommand{\isi}{\ensuremath{\mathit{ISI}}\xspace}

\newcommand{\gsiuc}{\ensuremath{\mathit{GSI_{\uc}}}\xspace}
\newcommand{\gsipd}{\ensuremath{\mathit{GSI_{\pd}}}\xspace}
\newcommand{\gsild}{\ensuremath{\mathit{GSI_{\ld}}}\xspace}
\newcommand{\gsiup}{\ensuremath{\mathit{GSI_{\up}}}\xspace}
\newcommand{\fsi}{\ensuremath{\mathit{FSI}}\xspace}

\newcommand{\parhead}[1]{\noindent{\bf {\em #1.}}}
\newcommand{\ie}[0]{{\em i.e.}}
\newcommand{\eg}[0]{{\em e.g.}}
\newcommand{\etal}[0]{{\em et al.}\xspace}
\newcommand{\footref}[1]{%
    $^{\ref{#1}}$%
}

\begin{document}

\title{\prometheus: \textsc{PRO}cedural \textsc{MET}hodology for
  developing \textsc{HE}uristics of \textsc{US}ability}

\author{Cristhy~Jiménez, Héctor~Allende~Cid,
  Ismael~Figueroa
  \thanks{The original version of this work appears in Spanish in the
    IEEE Latin American Transactions
    journal~\cite{jimenezAl:ieeelatam2017}. We present a version in
    English for further dissemination of our work. C. Jiménez,
    H. Allende Cid, I. Figueroa, Pontificia Universidad Católica de
    Valparaíso, Chile. \url{cristhy.jimenez.g@mail.pucv.cl},
    \url{hector.allende@pucv.cl}, \url{ismael.figueroa@pucv.cl}}}

\maketitle

\begin{abstract}
  Usability is a key discipline related to the development of modern
  software systems. Its goal is to assess the user-friendliness and
  effectiveness of a software product from the user point of
  view. Therefore, proper methodologies and techniques to perform this
  assessment are definitely relevant. Heuristic evaluation is probably
  the most commonly used method for usability assessment. Initially
  developed by Nielsen and Molich in the '90s, traditional heuristic
  evaluations rely on Nielsen's well-known 10 usability
  heuristics. However, recent evidence suggests that such heuristics
  are not sufficiently complete for dealing with new domains such as
  interactive television, virtual worlds, and many others. In addition
  to the lack of suitability of traditional heuristics, in the past
  years the lack of a robust methodology or process to effectively
  develop and validate these new domain-specific heuristics has been
  documented. In this paper we summarize current evidence regarding
  the lack of suitability of traditional heuristics, as well as the
  need to develop new domain-specific heuristics. After identifying
  and acknowledging existing gaps in heuristics for state-of-the-art
  technology, as documented by other researchers, we present
  \prometheus, a PROcedural METhodology for developing HEuristics of
  USability. \prometheus refines the methodology of Rusu ~\etal~
  (2011), and is composed of 8 stages. \prometheus clearly defines the
  artifacts that are required and produced by each stage, and also
  presents a set of quality indicators in order to assess the need for
  further refinement in the development of new heuristics. As an
  initial validation of \prometheus, we apply a questionnaire to
  several researchers that have used the methodology of Rusu ~\etal~,
  and we have also performed a small retrospective study, computing
  the quality indicators of several previous studies. Our results
  suggest that \prometheus is a very promising methodology, and that
  the metrics and indicators are indeed pertinent with respect to the
  conclusions of previous studies.
\end{abstract}


\section{Introduction}
\label{sec:introduction}

\IEEEPARstart{C}onstant technological development is reflected in the
development of new applications, many of them based on \emph{new
  information technologies} or \emph{emerging technologies}, such as
mobile devices, touch interfaces, distributed applications, virtual
learning environments, to name just a few. These applications
characteristically belong to a \emph{specific domain}, and are
accessed using \emph{different physical and logical devices}, as well
as \emph{specific contexts of use}. Since it is increasingly common
that non-specialist users need to access and use such applications on
a frequent basis, it is necessary and important to facilitate such
users’ experience to improve their satisfaction levels and task
performance.

\emph{Usability} is a discipline which, in the field of Software
Engineering, allows us to \emph{estimate the degree to which a
  software product can be used by specific users to achieve specific
  goals effectively, efficiently and satisfactorily in a specific
  context of use}~\cite{iso19989241}.
Therefore, one of its main activities is \emph{determining usability
  problems} in specific applications, in order to improve the quality
of this attribute in future iterations. Given the high costs of
usability \emph{tests}~\cite{UsabilityMethods}, where the behavior of
real users is observed and evaluated, two lower-cost techniques are
generally used to identify usability problems: \emph{the inquiry} and
the \emph{usability inspection}.
Usability inquiries characteristically use a qualitative approach
based on user opinions, and include methods such as questionnaires,
field studies, and different kinds of interview.
On the other hand, usability inspections consist of detailed
examinations carried out by \emph{groups of evaluators}, who report on
specific and general problems regarding a particular application.
One of the most widely used inspection methods is \emph{heuristic
  evaluations}, proposed by Nielsen and
Molich~\cite{NielsenMolich1990}. In this method, a small group of 3-5
evaluators inspect a particular software system's usability, guided by
a list of usability criteria known as \emph{usability
  heuristics}~\cite{NielsenMolich1990,luna2015using}. Heuristic
evaluation is a low-cost easily applicable method, which can be used
at various stages of an application’s
development~\cite{NielsenMolich1990}. It is also an effective method,
as it can detect around 75\% of usability problems with groups of 3 to
5 evaluators~\cite{Nielsen1993usability}. However, while heuristic
evaluations are traditionally carried out using Nielsen’s 10 usability
heuristics as criteria~\cite{JackobNielsen1995}, the need to develop
domain-specific heuristics, or put simply, \emph{domain heuristics},
is ever more pressing.

In a recent study, Hermawati and Lawson~\cite{hermawatiLawson:2016}
performed a bibliographic review of 70 articles related to the
development of domain heuristics, identifying two large problems:
significant deficiencies in efforts to validate new heuristics, and
lack of rigor, robustness and standardization in the effectiveness
analysis of domain heuristics. As a solution to these problems, in
this article we present \prometheus, a PROcedural METhodology for
developing HEuristics of USability, which refines the existing
methodology of Rusu~\etal~\cite{rusu2011methodology}--—which we call
\rrrc, based on its authors’ initials---for greater formality, rigor
and precision in the process of constructing new domain
heuristics. \prometheus has 8 stages, in each of which the activities
to be performed, the expected and produced artifacts, and the quality
indicators for heuristics — which in turn guide the 8 stages’
continuous refinement process — are described in detail. In this
article, we begin by presenting an overview of how domain heuristics
have been usually generated (\Cref{sec:how-doma-heur}) and the \rrrc
methodology (\Cref{sec:r3c-methodology}), to then present the main
contribution, which is \prometheus (\Cref{sec:prometheus}). Finally,
we conclude with the initial validation of \prometheus
(\Cref{sec:initial-validation}).

\section{How domain heuristics are developed}
\label{sec:how-doma-heur}

There is vast literature on heuristic evaluations and the development
of usability heuristics. In fact, as is studied by Hermawati and
Lawson~\cite{hermawatiLawson:2016}, there are at least 70 relevant
studies regarding the development of domain heuristics, so a detailed
review is beyond the scope of this article. Below we present a summary
of the main conclusions of~\cite{hermawatiLawson:2016}, which serve as
the foundation for the artifacts and stages proposed in \prometheus.

\subsection{Collecting and transforming information}
\label{sec:coll-transf-inform}

The first point to emphasize is that most of the studies analyzed in
\cite{hermawatiLawson:2016} include two large stages: collecting
information on heuristics, and transforming that information into
domain heuristics. Hermawati and Lawson describe 4 strategies for
collecting information:

\begin{enumerate}
\item Adopt existing theories.
\item Analyze context of use.
\item Analyze existing case study reports.
\item Analyze and/or create a common set of usability problems.
\end{enumerate}

Then, the transformation process uses one of the following 3
strategies:

\begin{enumerate}

\item Make a list of all the information, eliminating redundancies and
  overlaps, and using the end result as the new domain heuristics.

\item Perform the same normalization process, and sort the resulting
  information into categories, which are then transformed into
  heuristics.

\item Extend and/or modify Nielsen's heuristics.
  
\end{enumerate}

\subsection{Validating the heuristics}
\label{sec:valid-heur}

An important result from~\cite{hermawatiLawson:2016} is that they detected deficient
validation of domain heuristics. In short, 34\% of the analyzed
studies reported no validation. Of the remaining 66\%, the most
commonly used methods to validate domain heuristics are:

\begin{enumerate}
\item Experts applying the domain heuristics (24 studies).
\item Comparing the results with those from other heuristics (20
  studies).
\item Comparing the results, based on tests with real users (5
  studies).
\end{enumerate}

However, there are few studies that attempt to determine the
effectiveness of the new heuristics. As reported
in~\cite{hermawatiLawson:2016}, only 19 of the 70 cases studied
present a comparison in terms of effectiveness~\ie~the other cases do
not present validation, do not use a control group of heuristics, nor
do they perform any comparison or quantitative analysis. This
notwithstanding, in the cases where the new domain heuristics versus a
set of control heuristics are studied, the following categories of
validation techniques were identified:

\begin{enumerate}
\item Quantify all usability problems, per heuristic, and compare with
  an evaluation using control heuristics.

\item Quantify and compare the frequency, severity and distribution of
  those problems via each specific heuristic and via each set of
  heuristics.

\item Identify and quantify problems detected in both sets of
  heuristics, and unique problems detected separately by the
  heuristics sets.

\end{enumerate}

However, there is great variability in the extent and rigor of the
quantitative analyses. On the other hand, only 3 studies used the
predefined metrics proposed by
Hartson~\etal~\cite{hartsonAl:ijhci2003}. To conclude this summary,
the authors of~\cite{hermawatiLawson:2016} identified 3 areas to
improve the rigor and robustness of validating domain heuristics:

\begin{enumerate}

\item Adopt robust and rigorous validation metrics, to determine the
  effectiveness of new domain heuristics.
\item Create new domain heuristics based on existant heuristics in the
  domain.
\item Improve the definition and categorization of \emph{expert}, so
  as to control the variability introduced by evaluators.

\end{enumerate}

\section{\rrrc Methodology}
\label{sec:r3c-methodology}

In this section, and with the objective of contextualizing the
contribution of \prometheus, we describe the \rrrc methodology. This
is a methodology consisting of 6 stages, which we quote
textually~\cite{rusu2011methodology}:


\begin{enumerate}
\item \textbf{Exploratory}: to collect bibliography related with the
  main topics of the research: specific applications, their
  characteristics, general and/or related (if there are some)
  usability heuristics.

\item \textbf{Descriptive}: to highlight the most important
  characteristics of the previously collected information, in order to
  formalize the main concepts associated with the research.

\item \textbf{Correlational}: to identify the characteristics that the
  usability heuristics for specific applications should have, based on
  traditional heuristics and case studies analysis.

\item \textbf{Explicative}: to formally specify the set of the
  proposed heuristics, using a standard template.

\item \textbf{Validation (experimental)}: to check new heuristics
  against traditional heuristics by experiments, through heuristics
  evaluations performed on selected case studies, complemented by user
  tests.

\item \textbf{Refinement}: based on the feedback from the validation
  stage.
  
\end{enumerate}

\parhead{Evaluation criterion} \rrrc identifies and defines three
types of problem that can be detected in the experimental stage,
consisting of groups of evaluators who use the new heuristics, and
groups of evaluators using the control heuristics. Note that we use a
new notation to refer to these types of problem.

\begin{itemize}
\item \emph{Common problems}, \problemasComunes: Problems identified
  by both groups of evaluators.
\item \emph{Domain problems}, \problemasDominioUniq: Problems
  identified only by the group of evaluators who used the new
  heuristics.
\item \emph{Control problems}, \problemasControlUniq: Problems
  identified only by the group of evaluators who used Nielsen’s
  heuristics (or others if appropriate) as control.
\end{itemize}

With this sorting, \rrrc establishes that the new heuristics work well
when \problemasComunes and/or \problemasDominioUniq include the
highest percentage of problems. In practice, the severity of specific
and general problems is also considered. Finally, note that \rrrc is a
simple methodology, which has been applied successfully and offers a
framework that encourages experimental validation, thus avoiding some
of the problems described in~\Cref{sec:how-doma-heur}. It is precisely
because of this that we take \rrrc as starting point for the
development of \prometheus.

\section{\prometheus}
\label{sec:prometheus}

\prometheus is a \prometheuslemma, which has emerged as a refinement
of \rrrc. We say it is a \emph{procedural methodology} as it
accurately describes the steps to follow---and the artifacts to
construct---during the elaboration process of new domain
heuristics. \prometheus originates from a previous investigation by
Jiménez~\etal~~\cite{Jimenez2012SCCC} to determine the usability of
\rrrc. Therefore, they conducted a questionnaire with the researchers
who developed heuristics using \rrrc, in the domains of grid
computing~\cite{rusu2011usabilityGrid}, interactive
television~\cite{solano2011usability}, virtual
worlds~\cite{Munoz2012d} and touchscreen
devices~\cite{Inostroza2012}. On the quantitative side, the aggregated
results are inconclusive, because the ease of use was evaluated as
``neutral''. However, there were specific difficulties in the
explanatory, experimental and refinement stages. This was confirmed by
the comments made by those surveyed, which indicate lack of clarity in
these stages, and especially in refinement.

Next, we present \prometheus, starting with its contextualization
regarding \rrrc, and then describing the critical path for the
methodology. A full-fledged description of \prometheus is presented
in~\Cref{sec:deta-descr-prom}.

\begin{figure}[!t]
\centering
\includegraphics[scale=0.5]{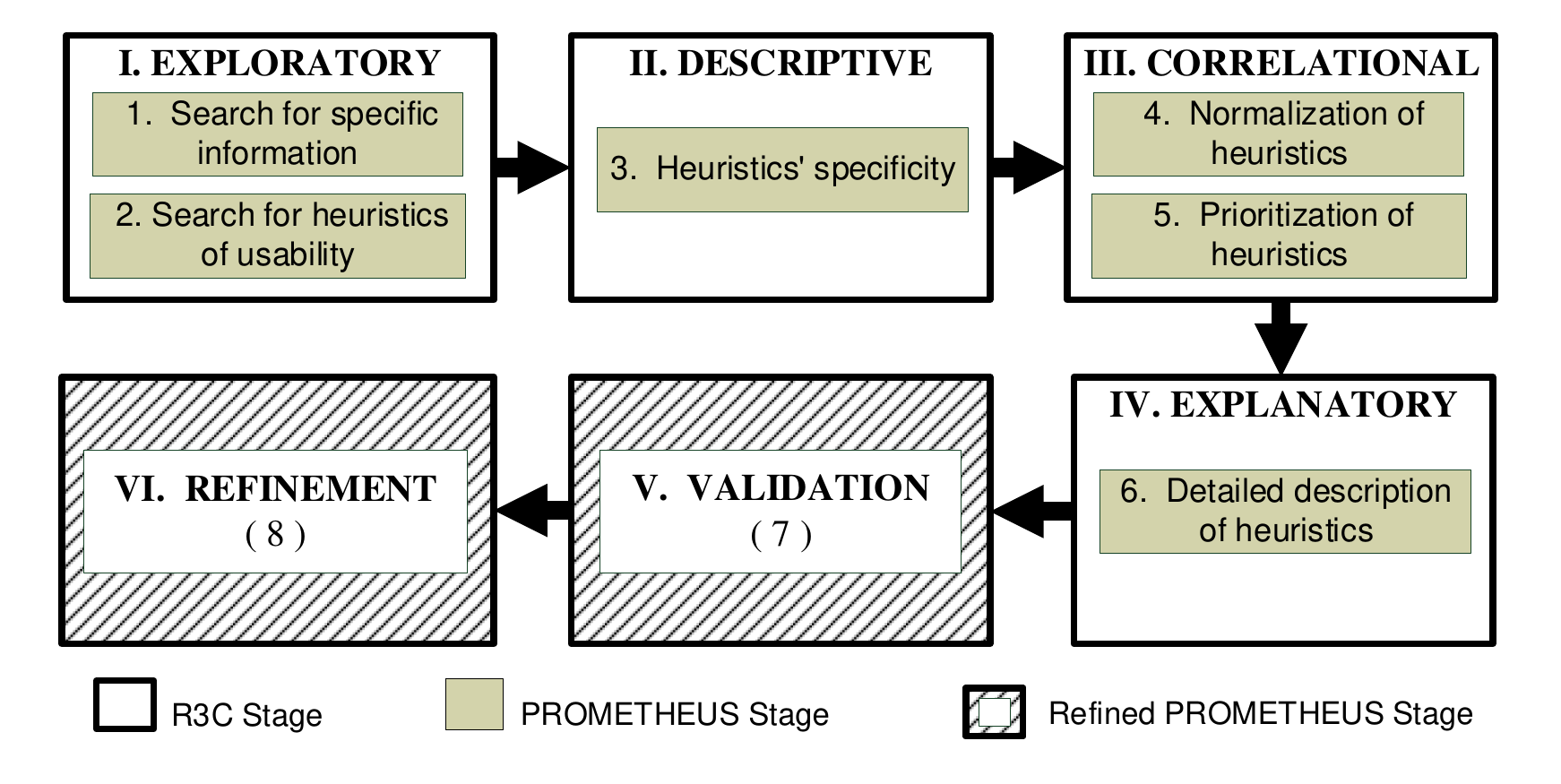}
\caption{Comparison between \rrrc stages and \prometheus stages.}
\label{fig:pdf_rusu_vs_prometheus}
\end{figure}

\subsection{Refinements to \rrrc}
\label{sec:refinements-r3c}

\Cref{fig:pdf_rusu_vs_prometheus} presents a diagram with the \rrrc
stages and the \prometheus stages. At a first glance, the main
difference is that \prometheus breaks up the exploratory and
correlational stages, each into two new stages. In addition, the
descriptive and explanatory stages have been renamed, and the
validation and refinement stages have been refined. The specific
contributions of \prometheus to the \rrrc stages are:

\begin{itemize}
\item \textbf{Exploratory}: \prometheus specifies the stages
  \emph{search for specific information} and \emph{search for
    heuristics of usability}. The first considers 4
  \emph{characteristic dimensions} to describe the specific domain:
  context of use, logical devices, physical devices, and user
  profiles. The second stage provides guides for a systematic
  literature review that produces a set of heuristics applicable to
  the domain.

\item \textbf{Descriptive}: \prometheus specifies an encoding table of
  \emph{specificity indices}, based on the domain’s characteristic
  dimensions.

\item \textbf{Correlational}: \prometheus proposes two stages:
  \emph{normalization of heuristics} and \emph{prioritization of
    heuristics}. In the first stage, cases of duplication or overlap
  found in the heuristics are resolved. The second case creates a
  specificity ranking for each heuristic, also considering the
  characteristic dimensions, among other factors.

\item \textbf{Explanatory}: the explanatory stage is kept from \rrrc,
  although recent evidence suggests it is possible to improve this
  stage~\cite{Jimenez_EAST2012}, \eg~considering novice evaluators.

\item \textbf{Validation}: we have kept the idea from \rrrc to perform
  experimental validation. However, \prometheus requires at least one
  heuristic evaluation using the domain heuristics and a group of
  control heuristics. In addition, \prometheus specifies quality
  indicators for the heuristics, based on this experiment.

\item \textbf{Refinement}: in \prometheus, the quality indicators
  obtained during the validation stage are used to suggest in which
  stages changes must be made or which specific problems need to be
  solved.

\end{itemize}

\begin{figure}[!ht]
\centering
\includegraphics[scale=0.68]{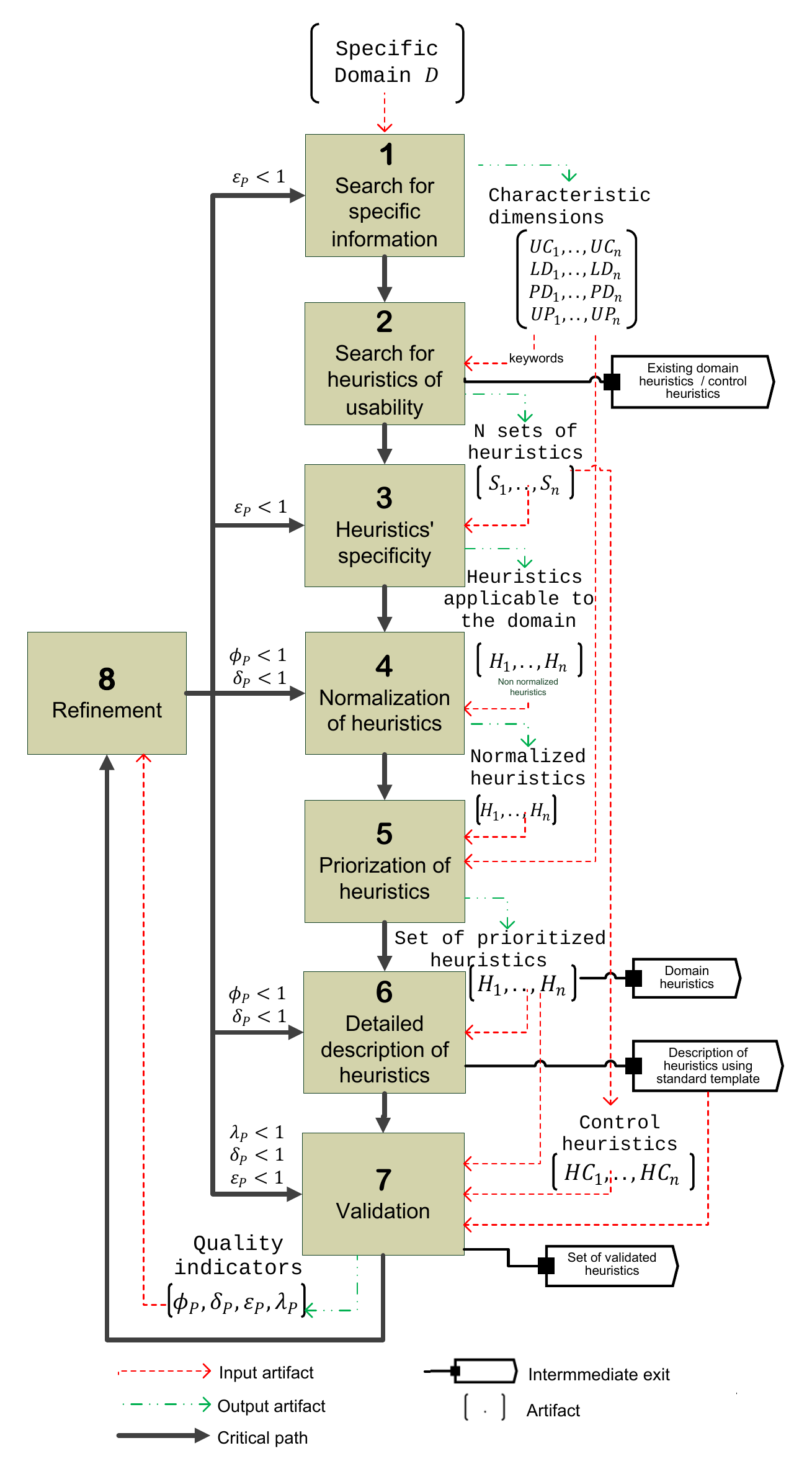}
\caption{Flow diagram of the main process of \prometheus.}
\label{fig:pdf_prometheus_flow}
\end{figure}

\subsection{Overview of the critical path}
\label{sec:overv-crit-path}

\Cref{fig:pdf_prometheus_flow} illustrates the 8 stages of \prometheus
to develop domain heuristics. The figure shows the required artifacts
as input, and the produced artifacts as output for each stage. It also
shows potential intermediate outputs, when suitable heuristics are
found, thus making the development of new heuristics
unnecessary. However, the methodology’s \emph{critical path} consists
of all the steps taken by a researcher for the effective development
of new usability heuristics for a specific domain. That is to say,
despite considering situations of intermediate outputs, \prometheus is
focused on situations where it is necessary to create new domain
heuristics.

For starters, in the first iterative application of the methodology,
all stages should be applied sequentially. Then, based on feedback
from the refinement stage, greater adaptability and flexibility are
possible in the stages performed during the progressive improvement of
the new heuristics. Obviously, the number of iterations and the stages
that have to be repeated will depend on the researcher's needs, the
results of the validation, and other factors specific to each
project. However, the indicators generated in the validation stage
allow us to make an informed decision whether or not to continue with
the refinement. The critical path is summed up in four major
phases:\newline

\subsubsection{Systematic search}
\label{sec:systematic-search}
The process starts by selecting the domain for which a set of specific
usability heuristics is required. Stages 1 to 5 consist in a
systematic search process, specificity encoding, and prioritizing one
or more sets of existing heuristics, based on the characteristic
domain dimensions, using a standard template.\newline

\subsubsection{Defining the domain heuristics}
\label{sec:defin-doma-heur}
At the end of stage 5, the researcher will have a set of heuristics
applicable to the domain, where there should be no duplication or
overlapping problems. Crucially, \emph{at this stage the researcher
  can create new heuristics} based on existing data. Then, stage 6
describes the heuristics using a standard template, like in \rrrc. The
detailed description aims to facilitate the evaluators’ task of
performing the experimental validation. The result of stage 5 and/or 6
is the new domain heuristics. \newline

\subsubsection{Experimental validation}
\label{sec:exper-valid}
Subsequently, stage 7 is to experimentally validate the domain
heuristics. Unlike \rrrc, and in line with the suggestions of
Hermawati and Lawson~\cite{hermawatiLawson:2016}, \prometheus
\emph{requires} at least one heuristic evaluation in a domain-specific
case study.
This evaluation \emph{must} consider the domain heuristics, and a set
of \emph{control heuristics}, which by default can be Nielsen's 10
heuristics, so as to generate the indicators that are detailed
below. It would also be ideal to have several groups of evaluators for
each set of heuristics, to minimize the differences between
groups. Another difference from \rrrc is that the evaluators that use
the control heuristic \emph{must} assign a specificity score to each
problem found concerning the domain. This score uses the same scale
used by the researcher in stage 3.

The validation phase culminates with the calculation of the following
indicators:

\begin{itemize}

\item \emph{Rate of unique problems},
  $\tasaUnicidad = \problemasDominioUniq / \problemasControlUniq$:
  This represents which group of heuristics found more unique
  problems. If $\tasaUnicidad > 1$, more problems were found in the
  domain heuristics.

\item \emph{Rate of problem dispersion}: To measure the distribution
  of problems in the groups of heuristics, we consider the values
  \dispersionControlUniq and \dispersionDominioUniq, which represent
  the standard deviation of problem distribution for the control
  heuristics and domain heuristics respectively. Given the above, we
  define the rate of dispersion
  $\tasaDispersion = \dispersionControlUniq /
  \dispersionDominioUniq$. If $\tasaDispersion > 1$, problems are
  better distributed in the domain heuristics than in the control.

\item \emph{Rate of severity}: Similar to the rate of dispersion, we
  define the rate of severity
  $\tasaSeveridad = \severidadDominioUniq / \severidadControlUniq$,
  which represents the relationship between the average severity
  \severidadDominioUniq of the problems encountered with the domain
  heuristics, versus the average severity \severidadControlUniq of the
  problems encountered with the control heuristics. If
  $\tasaSeveridad > 1$ it means that on average, the domain heuristics
  found more severe problems than the control heuristics.

\item \emph{Rate of specificity}: Finally, we define the rate of
  specificity
  $\tasaEspecificidad = \especificidadDominioUniq /
  \especificidadControlUniq$, which relates specificity averages
  \especificidadDominioUniq and \especificidadControlUniq, of the
  problems encountered with the domain and control heuristics
  respectively. If $\tasaEspecificidad > 1$, then the problems found
  by the domain heuristics are, on average, more specific that those
  found by the control heuristics.

\end{itemize}

It is important to remark that the indicators are uniform~\ie~when its
value is greater than 1, it is considered positive, and we consider
that there are potential efficiency problems when its value is less
than 1. This is useful, for example, to construct visualizations
like~\Cref{fig:indicadores}, which compares the indicator values from
various existing studies.

\subsubsection{Refinement}
\label{sec:refinement}

As shown in~\Cref{fig:pdf_prometheus_flow}, the refinement stage takes
the heuristic quality indicators \tasaUnicidad, \tasaDispersion,
\tasaEspecificidad and \tasaSeveridad into consideration, so as to
propose possible refinements to the domain heuristics. Such
refinements involve returning to work in one of the specific stages of
\prometheus, and then continuing with the critical path as needed. The
need for refinement was born, at least, from the following problematic
scenarios, or any combination of them.

\begin{itemize}

\item $\tasaUnicidad < 1$, that is, the domain heuristics found fewer
  unique problems. This may indicate that the heuristics are not well
  adjusted to the domain, that the application examined has mainly
  general problems, or else that the detailed explanations for the
  domain heuristics were not understood well by the evaluators.

\item $\tasaDispersion < 1$, that is, problems are worse distributed
  in the domain heuristics than in the control. This can be a symptom
  of overlapping heuristics, too many problems, or even overly
  specific heuristics, lack of problems, or that they are difficult to
  implement for evaluators.

\item $\tasaEspecificidad < 1 $, that is, the problems found by the
  control heuristics are considered more specific to the domain than
  those found by the domain heuristics. This problem can indicate
  problems with the prioritization of domain heuristics, or problems
  in the experiment design, for example selecting groups of
  evaluators.

\item $\tasaSeveridad < 1$, that is, the problems found by the control
  heuristics are more severe than those found by the domain
  heuristics. In general, this may indicate problems in the experiment
  design, in particular with the selection of the groups of
  evaluators.\newline

\end{itemize}

The end decision determining when the domain heuristics are considered
\emph{validated} depends ultimately on the researchers, the domain,
and the specific applications being studied. However, \prometheus
provides a precise methodological process with objective quality
indicators, which promote continuous feedback and provide quantitative
support to make such a decision.

\section{Initial Validation}
\label{sec:initial-validation}

For the initial validation of \prometheus we have used a qualitative
approach via a short questionnaire, taken by researchers who used
\rrrc in their work, and a quantitative approach by calculating
quality indicators (described in~\Cref{sec:overv-crit-path}) from the
existing results mentioned in~\Cref{sec:how-doma-heur}, as far as the
data allowed.

\subsection{Questionnaire for investigators that used \rrrc}
\label{sec:quest-invest-that}

We gave a questionnaire (see~\Cref{sec:interv-init-eval}) to 5
investigators that used \rrrc in cases studies related to domains of:
\emph{U-Learning}, applications for tablets, transactional web
applications, and smartphone applications. At the time the
questionnaire was taken, four of the interviewees were senior students
in Computer Engineering at the Pontificia Universidad Católica de
Valparaíso, and one, Inostroza, the author of~\cite{Inostroza201640},
was a doctoral student in the final stages of his thesis. Two
interviewees worked on tablet domains, but are considered as experts
individually, for the purposes of this validation. The objective of
the questionnaire was to have expert evaluation of the following
criteria:

\begin{itemize}

\item \emph{Objective of the study}
\item \emph{Applicable stages}
\item \emph{Stages that appear easiest to apply}
\item \emph{Stages that appear hardest to apply}
\item \emph{Recommendations for adding new stages}
\item \emph{Recommendations for eliminating stages}
\item \emph{General comprehension of \prometheus}
\item \emph{Specific comprehension of each stage}
\item \emph{Pertinence of quality indicators}
\item \emph{Ease of calculation of quality indicators}
\item \emph{Other recommendations}

\end{itemize}

We gave each participant a document with a preliminary description of
\prometheus, so they could evaluate the methodology’s applicability to
their case studies. We also carried out individual sessions to clarify
any doubts. After these sessions, the participants had three months to
evaluate the applicability of \prometheus.

\parhead{Results} \Cref{tab:resumenValidacion} summarizes the main
results from the questionnaires taken. The following information is
noteworthy:

\tymin=26pt
\newcommand{\na}{did not compute them}
\newcommand*{\thead}[1]{\textbf{#1}} 

\begin{table*}[!t]
  \begin{scriptsize}
  \centering
  \caption{Summary of the Initial Validation Results for \prometheus}
  \label{tab:resumenValidacion}
  \begin{tabulary}{\textwidth}{CLCCCCCCCCCC}

    \thead{Case Study}
    & \thead{Goal of the study}
    & \thead{Applicable}
    & \thead{Easiest}
    & \thead{Hardest}
    & \thead{Would add?} 
    & \thead{Would remove?}
    & \thead{Understands in general?}
    & \thead{Understands each stage?}
    & \thead{Useful?}
    & \thead{Indicators are pertinent?}
    & \thead{Indicators easy to compute?} \\
    \hline
    U-Learning 
    & Obtener nuevas heurísticas
    & 1, 2, 3, 4, 7, 8 
    & 1, 2, 3, 4, 7, 8 
    & 4
    & No 
    & No 
    & \multicolumn{1}{c}{\cmark} 
    & \multicolumn{1}{c}{\cmark} 
    & \multicolumn{1}{c}{\cmark} 
    & \multicolumn{1}{c}{\cmark} 
    & \multicolumn{1}{c}{\cmark} \\
    \hline
    Tablets \#1 
    & Obtain new heuristics, refine existing heuristics, describe
      heuristics in detail
    & 1, 2, 6, 7
    & 6, 7
    & -- 
    & No 
    & No 
    &  \multicolumn{1}{c}{\cmark} 
    & \multicolumn{1}{c}{\cmark} 
    & \multicolumn{1}{c}{\cmark} 
    & \na 
    & \na \\
    \hline
    Tablets \#2 
    & Obtain new heuristics, refine existing heuristics, describe
      heuristics in detail
    & 1, 2, 6, 7
    & 1, 2, 6, 7
    & -- 
    & No 
    & No 
    & \multicolumn{1}{c}{\cmark} 
    & \multicolumn{1}{c}{\cmark} 
    & \multicolumn{1}{c}{\cmark} 
    & \na 
    & \na \\
    \hline
    Transactional web applications
    & Refine existing heuristics, describe
      heuristics in detail
    & 6
    & 6
    & -- 
    & No 
    & No 
    & \multicolumn{1}{c}{\cmark} 
    & \multicolumn{1}{c}{\xmark} 
    & \multicolumn{1}{c}{\cmark} 
    & \na 
    & \na \\
    \hline
    Smartphones
    & Refine existing heuristics, describe
      heuristics in detail
    & 6
    & 6
    & -- 
    & No 
    & No 
    & \multicolumn{1}{c}{\cmark} 
    & \multicolumn{1}{c}{\cmark} 
    & \multicolumn{1}{c}{\cmark} 
    & \na 
    & \na \\
    \hline

\end{tabulary}  
\end{scriptsize}  
\end{table*}

\begin{itemize}

\item 3 of the 5 participants had the objective of constructing new
  heuristics, while the other 2 were looking to refine existing
  heuristics. This has an effect on the potential number of applicable
  stages, as the investigators from the first group considered more
  stages of \prometheus applicable to their projects. In the case of
  U-Learning, 6 of 8 stages were considered applicable. In contrast,
  for transactional web and smartphone applications, only stage 6 was
  considered applicable.
 
\item In general, participants considered that the applicable stages
  would also be easy to apply. The only explicit case of a stage found
  hard to apply is stage 4 (heuristic normalization), in the
  U-Learning domain.

\item All participants indicated they did not consider it necessary
  to add or eliminate stages within the methodology, that the general
  application of the methodology was clear and that the methodology
  was useful. Regarding comprehension of each particular stage, there
  were only issues with understanding stage 6 in the case of
  transac-tional web applications.

\item Unfortunately, the quality indicators were only consid-ered
  applicable, pertinent and easy to calculate in the case of
  U-Learning. In general, the other investigators were not interested
  in calculating them.

\item Among the recommendations given, the following stand out:

  \begin{itemize}
  \item Include an example of how to fill the template in for the
    ``detailed heuristic description'' stage.

  \item Better explain the calculations included in the methodology.

  \item Include a real application example.
  \end{itemize}

\end{itemize}

\subsection{Validation of quality indicators}
\label{sec:valid-qual-indic}

\parhead{Preliminaries} Before showing the obtained results, let us
recall the classification of problems proposed in \rrrc
(\Cref{sec:r3c-methodology}):

\begin{itemize}
\item \emph{Common problems}, \problemasComunes: Common problems
  identified by both groups of evaluators.
\item \emph{Domain problems}, \problemasDominioUniq: Problems only
  identified by the group of evaluators that used the new heuristics.
\item \emph{Control problems}, \problemasControlUniq: Problems only
  identified by the group of evaluators that used Nielsen’s heuristics
  (or oth-ers if applicable) as control.
\end{itemize}

Now, let \setProblemas be the total set of problems found when
applying both the control heuristics and the domain heuristics; we
have that:

\begin{align*}
\problemasDominio &= \problemasDominioUniq \cup \problemasComunes
                    \text{~--~(domain heuristics problems)}\\
\problemasControl &= \problemasControlUniq \cup \problemasComunes
                    \text{~--~(control heuristics problems)}\\
\setProblemas &= \problemasDominioUniq \cup \problemasControlUniq \cup
                \problemasComunes = \problemasDominio \cup \problemasControl \text{~--~(total problems)}\\
\end{align*}

Similarly, in addition we consider the average severity
\severidadDominio of the problems in \problemasDominio and the average
severity \severidadControl of the problems in
\problemasControl. Considering all of the aforementioned, we define
the indicators
$\tasaUnicidadAprox = \problemasDominio / \problemasControl$ and
$\tasaSeveridadAprox = \severidadDominio / \severidadControl$ that
approximate the uniqueness and severity rates, respectively, in cases
where the common problems \problemasComunes are not separated between
the two groups of heuristics.\newline

Considering the previous definitions, the second point of the initial
\prometheus validation consists in analyzing the data from the 19
studies identified by~\cite{hermawatiLawson:2016} that attempt to
determine the effectiveness of the new heuristics
(\Cref{sec:how-doma-heur}), and considering two additional
recently-published
studies~\cite{luna2015using,carrare2015usability}. We calculate the
heuristic quality indicators for each of them as possible according to
the published information, and we underscore the conclusions regarding
whether or not the developed heuristics are more effective than the
control heuristics. \Cref{tab:resumenDSH} summarizes the methodology,
the criteria used by the authors, the indicators that could be
calculated, and whether or not the authors conclude that the domain
heuristics developed are more effective than the control heuristics
used. Here we synthesize the main results obtained. \newline

\parhead{Regarding the number of problems}
Of the 21 studies analyzed, we were able to calculate an indicator for
17 of them. In these 17 cases, the most commonly used indicator
corresponds to the uniqueness rate \tasaUnicidad, used in 9 cases, or
the approximate uniqueness rate \tasaUnicidadAprox, used in 4
cases. Two other cases only considered the number of problems based on
the domain heuristics; in other words, only \problemasDominio was
considered. In the remaining two cases, no indicator based on the
number of detected problems could be calculated.

\parhead{Regarding the dispersion of problems}
In a very distant second place, calculating the dispersion rate
\tasaDispersion was possible in four cases, while two cases only had
information regarding the problem dispersion for the domain
heuristics; in other words, $\sigma(\problemasDominio)$.

\parhead{Regarding the severity and specificity of the problems}
Similarly, it was possible to calculate the severity rate
\tasaSeveridad in four cases. In 2 cases it was possible to calculate
an approximation \tasaSeveridadAprox, and in another 2 cases only the
problem severity of the domain heuristics,~\ie~ \severidadDominio, was
considered.  Moreover, it was not possible to calculate the problem
specificity rate \tasaEspecificidad.

\parhead{Domain heuristics vs control heuristics}
In 20 of the 21 cases, the authors conclude that the domain heuristics
developed (or refined) are more efficient than then control
heuristics. The case wherein this does not occur is in the domain of
security management applications~\cite{Jaferian2011}; in this case
only $\tasaUnicidad = 1.03$ could be calculated, which is difficult to
interpret without additional information.

\begin{figure}[t!]
  \centering
  \includegraphics[width=1.0\linewidth]{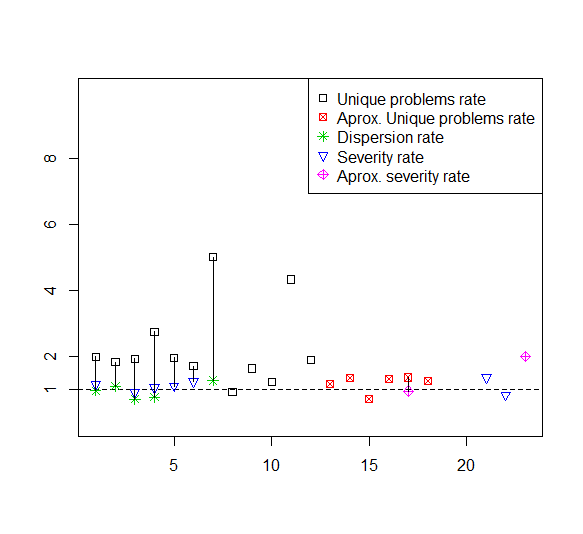}
  \caption{Visualization of quality indicators in selected studies
    where they could be calculated. Each sub-case is considered
    individually, thus 23 cases in total. In each case the authors
    conclude that the domain heuristics are better than those the
    control. An extreme case with $\tasaUnicidad = 9.1$ has been
    omitted. The dotted line indicates the value $1$, which
    differentiates between a positive or negative interpretation of
    the indicator.}
  \label{fig:indicadores}
\end{figure}

\parhead{Relation between indicators and positive conclusions}
\Cref{fig:indicadores} shows the indicators calculated
in~\Cref{tab:resumenDSH}, breaking each sub-case down as an
independent input in the graph. Each vertical line represents a case
study involving more than one indicator. A fundamental observation is
that the condition $\tasaUnicidad > 1$, or its approximation
$\tasaUnicidadAprox > 1$, seems to be the best predictor to validate
the domain heuristics’ efficiency. This is not so in only two cases,
for $\tasaUnicidad = 0.93$ and $\tasaUnicidadAprox = 0.73$. Under the
\prometheus scheme, these cases are difficult to interpret and would
require a refinement, as well as calculating the other specified
indicators. Furthermore, other qualitative factors may contribute to
positive evaluation in these case studies. Regarding the distribution
and severity rates, we observed values close to $1$, and which
complement the uniqueness rates. In just 3 cases, only severity was
considered, due to a lack of other quantitative data.

To conclude this section, we observe that the retrospective analysis
of the selected cases in which domain heuristics have been developed
supports the pertinence of the indicators proposed in
\prometheus. Particularly, it is clear that the most important
indicator is the uniqueness rate, and that the other indicators are
more complementary and can help discriminate in less conclusive
cases. We also observed existent studies that base their positive
conclusions on a rather weak quantitative argument, which suggests
that the refinement recommendations in \prometheus are on the right
track.

\section{Conclusions and Future Work}
\label{sec:concl-future-stud}

\prometheus is a \prometheuslemma proposed to increase the robustness
of investigations that aim to develop new domain-specific usability
heuristics. Based on Hermawati y Lawson’s
study~\cite{hermawatiLawson:2016}, and using the \rrrc methodology as
a base~\cite{rusu2011methodology}, we propose a process consisting in
7 principal stages, and a continuous refinement cycle. The main
contributions of \prometheus’ are: (i) it specifies each artifact
consumed or generated in detail for each stage, (ii) it defines
quantitative quality indicators, and (iii) it defines the refinement
based on feedback provided by the indicators. Based on a questionnaire
completed by expert investigators, and a retrospective analysis on
relevant, state-of-the-art case studies, we have performed an initial
validation on \prometheus which is, in our opinion, very positive and
suggests new and immediate lines of future work for the additional and
complete validation of \prometheus.

In essence, it is necessary to apply \prometheus using real case
studies that aim to develop new domain heuristics, and follow the
critical path, so as to obtain specific feedback in each of the
proposed stages. To this end, we propose applying \prometheus to two
new case studies:

\begin{enumerate}
\item \emph{Virtual Learning Environments}, with the case study in
  question to be found at http://elearning.espoch.edu.ec.
\item E-Government Pages. Also based on public ser-vices available in
  Ecuador.
\end{enumerate}

Both of these cases are expected to have at least two groups of
evaluators in the experimental validation stage. In addition,
semi-structured interviews and questionnaires will be conducted in
order to obtain the opinion of everyone involved in these case
studies.

{
\hyphenation{in-for-ma-ción smart-pho-nes geo-grá-fi-ca pre-sen-ta-ción}
\onecolumn
\begin{center}
\begin{scriptsize}
\begin{longtable}{@{\extracolsep{\fill}}|p{2.2cm}|p{0.5cm}|p{3.5cm}|p{3cm}|p{3.8cm}|p{2cm}|@{}}

\caption{Domain Heuristics \heuristicasDominio vs Control Heuristics
  \heuristicasControl - Efficiency Analysis Summary} \label{tab:resumenDSH} \\

    \hline
    \multicolumn{1}{|c}{Domain} & 
    \multicolumn{1}{c}{\# Hs} &
    \multicolumn{1}{c}{Criteria} &
    \multicolumn{1}{c}{Methodology} &
    \multicolumn{1}{c}{Indicators} &
    \multicolumn{1}{c|}{\heuristicasDominio better than \heuristicasControl?} \\
    \endfirsthead

    \hline
    \multicolumn{1}{|c}{Domain} &
    \multicolumn{1}{c}{\# Hs} &
    \multicolumn{1}{c}{Criteria}  &
    \multicolumn{1}{c}{Methodology} &
    \multicolumn{1}{c}{Indicators} &
    \multicolumn{1}{c|}{\heuristicasDominio better than \heuristicasControl?} \\
    \endhead

    \hline \multicolumn{6}{|r|}{{Continues in the next page...}} \\ \hline
    \endfoot

    \endlastfoot

    \hline 
    Grid Computing~\cite{rusu2011usabilityGrid}
    & 12
    & Quantity of unique, specific and general problems. Severity of
      specific vs general problems
    & \rrrc
    & {\textbf{Case 1:}
      $\tasaUnicidad = 2.0, \tasaDispersion = 0.97, \tasaSeveridad = 1.16$
      \textbf{\newline Case 2:} $\tasaUnicidad = 1.83, \tasaDispersion =
      1.11$ 
    }
    & \cmark \\

    \hline Virtual worlds~\cite{Munoz2012d}
    & 16
    & Quantity of unique, specific and general problems. Severity of
      specific vs general problems & \rrrc & \textbf{Case 1:}
    $\tasaUnicidad = 1.93, \tasaDispersion = 0.71, \tasaSeveridad
    =0.91$
    \textbf{\newline Case 2:}
    $\tasaUnicidad = 2.75, \tasaDispersion = 0.78, \tasaSeveridad
    =1.07$ & \cmark \\

  \hline Inter-cultural web sites~\cite{diaz2013cultural}
    & 13
    & Quantity of unique, specific and general problems. Severity of specific
      vs general problems &\rrrc
    & \textbf{Case 1:}
      $\tasaUnicidad = 1.97, \tasaSeveridad = 1.11$ \newline
      \textbf{Case 2:} $\tasaUnicidad = 1.71, \tasaSeveridad =
      1.25$\newline(\emph{based on percentages}) & \cmark \\

    \hline Digital Television~\cite{solano2013evaluating} & 14 &
    Quantity of unique, specific and general problems. Severity of specific
      vs general problems & \rrrc &
    \newline $\tasaUnicidad = 5.0, \tasaDispersion = 1.28$ & \cmark \\

    \hline Notification systems~\cite{Berry2003} & 8 & Quantity of
                                                       problems found
                                                       on 3 interfaces
                                                       used as case
                                                       study. Level of
                                                       compliance of
                                                       the heuristics
                                                       developed
    & Based on usability problems related to notification systems
    &
    \newline \textbf{Case 1:} $\tasaUnicidad = 0.93$
    \newline \textbf{Case 2:} $\tasaUnicidad = 1.64$
    \newline \textbf{Case 3:} $\tasaUnicidad = 1.23$
    & \cmark \\

    \hline Asistential Robotics~\cite{tsui2010developing} & 9 &
    Quantity of total and unique problems & Adaptation of existing
                                            heuristics & \newline $\tasaUnicidad = 4.33$ & \cmark \\

    \hline Security management applications~\cite{Jaferian2011} &
    7 & Quantity and severity of problems, effectiveness,
        meticulousness, validity and reliability of the heuristics &
                                                                     It
                                                                     uses
                                                                     \emph{grounded
      theory analysis} to generate the heuristics &
    \newline $\tasaUnicidad = 1.03$ & \xmark \\

    \hline Educational web sites~\cite{Alroobaea2013} & n/a &
    Quantity and severity of problems, effectiveness, meticulousness,
    validity and reliability of the heuristics, time used and associated cost
    & Own 4 steps methodology that analyzes the characteristics of the
      domain and evaluates cases through experts and users & \newline $\tasaUnicidad = 9.17$ & \cmark \\

    \hline Mobile-based applications~\cite{Neto2013} & 11 &
    Quantity and severity of problems & Extends Nielsen's heuristics & \newline $\tasaUnicidad = 1.9$ & \cmark \\

  \hline Mobile computing~\cite{Bertini2006} & 8 &
   Quantity and severity of problems, quantity of problems per
   heuristic, time used by evaluators & Own 3 steps methodology based
   on the analysis of 3 experts &
    \textbf{Case 1:} $\problemasDominio = 26,
    \problemasControl = 22$\newline$\tasaUnicidadAprox = 1.18$
    \newline \textbf{Case 2:} $\problemasDominio = 38,
    \problemasControl = 28$\newline$\tasaUnicidadAprox = 1.36$
    \newline \textbf{Combined cases:} $\tasaDispersion = 0.57$
    & \cmark \\

    \hline Learning based in clinic cases~\cite{carrare2015usability}
                                & 22 & Quantity and severity of
                                       problems in specific vs general
                                       heuristics & It uses
    \emph{grounded theory analysis} to generate the heuristics &
    \textbf{Case 1:} $\problemasDominio = 27, \problemasControl =
    37$\newline $\tasaUnicidadAprox = 0.73, \severidadDominio < \severidadControl$ \newline
    \textbf{Case 2:} $\problemasDominio = 74, \problemasControl =
    56$\newline $\tasaUnicidadAprox = 1.32, \severidadDominio > \severidadControl$
    & \cmark \\

    \hline Online games~\cite{Pinelle2009} & 10 &
    Quantity and severity of problems, quantity and severity of
                                                  problems per
                                                  heuristic
    & Based on the usability analysis on games &
    $\problemasDominio = 67, \problemasControl=49, \tasaUnicidadAprox = 1.37$
    \newline $\severidadDominio = 2.30, \severidadControl = 2.39, \tasaSeveridadAprox = 0.96$
    \newline $\sigma(\problemasDominio) = 3.16$
    & \cmark \\

    \hline Mobile map applications~\cite{Kuparinen2013} & 10 &
    Quantity and severity of problems & Adaptation of Nielsen's
                                        heuristics using a
                                        theoretical-conceptual focus &
    \newline $\problemasDominio = 19, \problemasControl = 15,
    \tasaUnicidadAprox = 1.27$
    & \cmark \\

    \hline Smartphones~\cite{Inostroza201640} & 12 & Quantity of
                                                     unique, specific
                                                     and general
                                                     problems. Severity
                                                     of specific vs
                                                     general problems & \rrrc & \newline
    $\mathit{\overline{\problemasDominio~~= 28.33}}, \mathit{\overline{\severidadDominio}=2.36}$\newline
    $\mathit{\overline{\sigma(\problemasDominio) = 2.24}}$\newline 
    & \cmark \\

    \hline Groupware~\cite{luna2015using}\footref{NoControl} & 24 &
    Problems found in case study and their severity & Based on design patterns &
    $\problemasDominio = 39, \sigma(\problemasDominio) = 1.85$ \newline
    $\severidadDominio = 2.15$
    & \cmark~Without \heuristicasControl, positive conclusions \\

    \hline Mobile touch devices~\cite{Inostroza2012_JCC} & 12
    & Quantity of unique, specific and general problems. Severity of
      specific vs general problems & \rrrc & \textbf{Case 1:}
    $\tasaSeveridad = 1.37$ \newline \textbf{Case 2:}
    $\tasaSeveridad = 0.83$ & \cmark \\

    \hline ERP Applications~\cite{Singh2009} & 5 & Mean, median and
                                                   mode of problem
                                                   severity per
                                                   heuristic & Based
                                                               on 5
                                                               evaluation
                                                               criteria
                                                               for ERP
                                    & \newline
    $ \severidadDominio = 1.76$;
    $ \severidadControl = 0.88, \tasaSeveridadAprox = 2$ \newline
    & \cmark \\

  \hline Ambient display~\cite{Mankoff2003} & 12 & Percentage and
                                                   severity of
                                                   problems identified
                                                   by both groups of heuristics
                                                   Porcentaje y
   & Adaptation of Nielsen's heuristics taking into account
     characteristics of the domain and the opinion of experts and
     designers
   & \newline The presented results do not permit us to compute the indicators & \cmark \\

    \hline Computer games~\cite{desurvire2009game}\footnote{\label{NoControl} No
  comparison against control heuristics} & 18
   & Kinds of problems, quantity and severity & Based on the
                                                literature and the
                                                review of experts in
                                                gameability and game
                                                designers  & The
                                                             presented
                                                             results
                                                             do not
                                                             permit us
                                                             to
                                                             compute
                                                             the
                                                             indicators
                                   &
    \cmark \\

    \hline Information display in big screens~\cite{Somervell2005} & 8 & Quantity of problems, quantity
                                       of real problems,
                                       meticulousness, validty,
                                       effectiveness and reliability
                                       of the heuristics
   & It uses critical parameters using scenario-based designs &
                                                                \newline
                                                                The
                                                                presented
                                                                results
                                                                do not
                                                                permit
                                                                us to
                                                                compute
                                                                the indicators & \cmark \\

    \hline Web Sites~\cite{Conte2009} & 13 & Quantity of problems,
                                             time used by evaluators
    & Adaptation of Nielsen's heuristics
    & The presented results do not permit us to compute the indicators & \cmark \\

    \hline

\end{longtable}
\end{scriptsize}
\end{center}
\twocolumn
}

\bibliographystyle{IEEEtran}


\clearpage
\appendices

\section{Detailed Description of \prometheus}
\label{sec:deta-descr-prom}

In this section we present a detailed explanation of each of the 8
stages of \prometheus. We describe the purpose of each stage, specify
input and output artifacts, and discuss their relevance. As shown in
Figure 2, the eight stages of \prometheus are:

\begin{enumerate}
\item Search for specific information
\item Search for usability heuristics
\item Heuristic specificity
\item Heuristic normalization
\item Heuristic prioritization 
\item Detailed description of heuristics
\item Validation 
\item Refinement
\end{enumerate}

\subsection{Stage 1: Search for specific information}
\label{sec:stage-1:-search}

The first stage is exploratory in nature, and its main objective is to
determine the characteristic dimensions of a specific domain,
\dom. The characteristic dimensions proposed are the following:

\begin{enumerate}
\item Usage contexts, \uc.
\item Interactive logic devices, \ld.
\item Interactive physical devices, \pd.
\item User Profiles, \up.
\end{enumerate}

\parhead{Input} The name or description of domain \dom is required to
begin to determine the characteristic dimensions. For example: ``grid
computing'', ``e-learning'', etc.

\parhead{Output} This step produces two artifacts. The first is a list
of relevant \emph{keywords} for the bibliographic search in the next
step (\Cref{sec:stage-2:-search}). The second is the list of
characteristic dimensions and their \emph{initial specificity
  indices}, which is summarized in tables, such as~\Cref{tab:uc}, one
for each characteristic dimension. The specificity values follow a
standard 5-level Likert scale, where 0 is not specific and 4 is
completely specific. These specificity scores are used in step 5
(\Cref{sec:stage-5:-prior}) for calculating the final specificity
index.

\begin{table}[h!]
  \centering
  \begin{tabular}{ll}
    \thead{Usage Context} & \thead{Specificity} \\
    \hline
    Indoor  & 3    \\
    Outdoor & 2    \\
    Noisy & 2    \\
    Quiet & 4 \\ 
    \hline    
  \end{tabular}
  \caption{Initial Specificity Example Table for Usage Context}
  \label{tab:uc}
\end{table}

An immediate goal for future work is to make a list of pre-selected
choices for each characteristic dimension. For example; suggest
physical devices such as desktops, tablets, phones, touchscreen, etc.,
to help complete this stage.


\subsection{Stage 2: Search for usability heuristics}
\label{sec:stage-2:-search}

The objective of this stage is to conduct a bibliographical search to
identify sets of usability heuristics related to the domain \dom under
study. In general, and as suggested research practice, it is
recommendable to perform a \emph{systematic mapping} or a
\emph{systematic review} of
literature~\cite{petersenAl:ease2008,kitchenham:east2012}.

\parhead{Input} At this stage the \emph{keywords} related to the
domain are used, from Stage 1 (\Cref{sec:stage-1:-search}).

\parhead{Output} At this stage sets of heuristics are obtained
$S_1, \dots, S_n$, which are potentially applicable to the domain. A
unique identifier must be assigned to each heuristic, for example,
$H_1^{S_1}$ represents the first heuristic of the set $S_1$.
If there is a set of heuristics validated for the domain and which
suits the researchers’ needs, it is possible to terminate the
application of \prometheus at this point. These heuristics can also
serve as control heuristic in Stage 7 (\Cref{sec:stages-7-8}), in the
event that new heuristics for the domain are defined.\newline

In general, Stages 1 and 2 specify how to carry out the process of
\emph{extracting information}, as reported by Hermawati and Lawson
\cite{hermawatiLawson:2016}, giving specific recommendations on how to
conduct the search, and taking into account the determining aspects of
the domain under study. As mentioned in \Cref{sec:concl-future-stud},
and in this appendix, there is considerable potential for improvement
in these stages.


\subsection{Stage 3: Heuristic specificity}
\label{sec:stage-3:-heuristic}

In order to perform an initial filtering, at this stage we assign an
\emph{initial specificity index} \isi, associated with each heuristic
found in the previous stage.

\parhead{Input} Heuristic sets $S_1, \dots, S_n$, identified in the
previous stage, are required.

\parhead{Output} At this stage a tabulation of heuristics is obtained
with their initial specificity indices, as shown
in~\Cref{tab:isis}. We say that these heuristics are
\emph{denormalized}, since they don’t resolve duplication and/or
overlap conflicts between the heuristics yet. However, it is possible
to rank the heuristics and eliminate those considered unlikely to be
applied to the domain.

\begin{table}[h!]
  \centering
  \begin{tabular}{ll}
    \thead{Heuristic} & \thead{ISI} \\
    \hline
    \newline & \\
    $H^{S_1}_1$  & 3      \\
    ~~$\vdots$    & $\vdots$      \\
    $H^{S_1}_{k_1}$ & 4    \\ 
    ~~$\vdots$ & $\vdots$  \\
    $H^{S_n}_{k_n}$ & 5    \\
    \newline & \\
    \hline    
  \end{tabular}
  \caption{Specificity Indices for Denormalized Heuristics}
  \label{tab:isis}
\end{table}


\subsection{Stage 4: Heuristic normalization}
\label{sec:stage-4:-heuristic}

The objective of this stage is to resolve possible cases of overlap
or duplication in the heuristics obtained so far.

\parhead{Input} The denormalized heuristics are received as input,
along with their initial specificity indicators.

\parhead{Output} At this stage a preliminary set of heuristics
applicable to the domain is produced, ensuring that there are no
overlap-ping or duplication problems between heuristics. Each
heuristic will have an \isi score, possibly revised after
normalization.

In general, cases of multiplicity can be solved through any of the
following:

\begin{itemize}
\item Keep one of the heuristics that have similarity and discard the
  others.

\item Discard similar heuristics and reformulate a new heu-ristic that
  combines the characteristics of the heuristics that make it up.
\end{itemize}

\noindent
Thus, cases of overlapping can be solved by any of the fol-lowing
strategies:

\begin{itemize}
\item Maintain a general heuristic that groups together the
  overlapping heuristics.

\item Separate the heuristic overlapping into several individual
  heuristics.
\end{itemize}

This process of resolving overlap and multiplicity should be carried
out iteratively until obtaining a set of normalized heuristics, in
other words, a set without any overlaps or multiplicities. Note that
if new heuristics are created, following the aforementioned
strategies, these must also have a unique identifier. Finally, it is
recommended to review and reconsider the values of the \isi
specificity for each one the normalized heuristics.


\subsection{Stage 5: Prioritization of Heuristics}
\label{sec:stage-5:-prior}

The objective of this stage is to synthesize the applicability of the
normalized heuristics, considering the specificity of each one with
respect to context of use, as well as the initial specificity
indicators, to have a rankable list of heuristics that can be applied
to the domain.

\parhead{Input} This stage receives the normalized heuristics obtained
in the previous stage, together with the respective \isi
indicators. In addition, it works with characteristic dimensions, \uc,
\pd, \ld, \up; obtained in Stage 1.

\parhead{Output} The principal product for this stage is the
\emph{specificity matrix}, as per the example
in~\Cref{tab:matrizEspecificidad}, which facilitates a final process
of ranking and selecting the heuristics that can be considered most
applicable for the domain.

\begin{table}[h!]
\begin{center}
\caption{Example of Specificity Matrix for Heuristics}
\begin{tabular}{|l|l|l|l|l|l|l|}
\hline
Heuristic & \isi & \gsiuc & \gsipd & \gsild & \gsiup & \fsi \\
\hline
$H^{S_3}_2$ & 3 & 4  & 2  & 1  & 3 & 1.875 \\
\hline
$H^{S_3}_3$ & 1 & 1 & 2  & 4 & 0 & 0.4375 \\
\hline
$H^{S_4}_2$ & 1 & 3  & 2 & 3  & 0 & 0.5 \\
\hline
$\vdots$ & $\vdots$ & $\vdots$  & $\vdots$  & $\vdots$ & $\vdots$ & $\vdots$\\
\hline
\end{tabular}

\label{tab:matrizEspecificidad}
\end{center}
\end{table}

\noindent
This matrix contains, for each heuristic $H_j$, a summary of the
following specificity indicators:

\begin{itemize}
\item \isi: initial specificity indicators, created in Stage 1 and
  refined in Stage 4.

\item \gsiuc, \gsipd, \gsild and \gsiup: global specificity indicators
  of each characteristic dimension. To exemplify the calculation of
  these indicators, we consider the \uc dimension, for which we
  created a tabulation such as~\Cref{tab:GSIUC}. For each usage
  context, a specificity score was assigned, and the \gsiuc score is
  the average of each row. Similarly, tables were created for the
  other characteristic dimensions.

\begin{table}[h!]
\begin{center}
\caption{Specificity for Context of Use and \gsiuc calculation}
\begin{tabular}{|l|l|l|l|l|l|}
\hline
Heuristic & Indoor & Outdoor & Noisy & Quiet & \gsiuc \\
\hline
$H^{S_3}_2$ & 4 & 4 & 4 & 4 & 4 \\
\hline
$H^{S_3}_3$ & 4 & 0 & 0 & 0 & 1 \\
\hline
$H^{S_2}_4$ & 4 & 2 & 2 & 4 & 3 \\
\hline
$\vdots$ & $\vdots$ & $\vdots$ & $\vdots$  & $\vdots$ & \\
\hline
\end{tabular}

\label{tab:GSIUC}
\end{center}
\end{table}

\end{itemize}

Finally, the final specificity index \fsi synthesizes the specificity
and applicability of the different heuristics to the domain in a
single value, considering the initial evaluation and the evaluation of
each characteristic dimension.
For a heuristic $H_j$, the index is calculated with the following
formula:

\[ \fsi_{H_j} = 4 * \frac{\isi_{H_j} * \sum \mathit{GSI}_{\{\uc, \ld, \pd, \up\}}}{64} \]

\noindent
The final value of \fsi varies between 0 and 4; in general, a high
value of \fsi is expected to indicate high applicability of that
heuristic, although the final selection criterion is always within the
discretion of the researchers involved, with the methodological
backing granted by \prometheus

\subsection{Stage 6: Detailed description of heuristics}
\label{sec:stage-6:-detailed}

At this stage the description of the heuristics selected in the
previous stage are formalized, in order to design the experimental
validation for Stage 7. This formal description can provide
information necessary to understand and apply the domain heuristics
when performing a heuristic evaluation, particularly for inexperienced
evaluators, as evidenced
in~\cite{Jimenez_EAST2012}. \Cref{tab:plantillaEstandar} presents a
template format that can be used at this stage. This template takes
the basis proposed by \rrrc~\cite{rusu2011methodology}, and adds a new
field, \emph{Checklist}, with the intention to further facilitate its
application, following the suggestions in~\cite{Jimenez_EAST2012}.

\parhead{Input} The selected set of heuristics for application to the
domain.

\parhead{Output} The formal description of each heuristic, based on
the standard template.

\begin{table}[!h]
\begin{center}
\caption{Standard Template to Describe Usability Heuristics.}
\begin{tabular}{| c  | p{5cm}  |}
  \hline
  \multicolumn{2}{|c|}{Heuristic Identifier}\\
  \hline
  Name & Name that identifies the heuristic. \\
  \hline
  Description & Detailed explanation of the heuristic. \\
  \hline
  Examples & Examples of compliance and non-compliance related to the
             heuristic. \\
  \hline
  Benefits & Expected usability benefits when there is compliance with
             the heuristic. \\
  \hline
  Problems & Expected problems that can arise if the heuristic is
             misunderstood during heuristic evaluation. \\
  \hline
  Application context & Additional information regarding the
                        applicability of the heuristic. \\
  \hline
  Related heuristics & References to other heuristics related to the
                       (non-)compliance of this heuristic. \\
  \hline
  Checklist & Detailed operational steps and criteria to be used when applying
              this heuristic. \\
  \hline
\end{tabular}

\label{tab:plantillaEstandar}
\end{center}
\end{table}


\subsection{Stages 7 \& 8: Validation and Refinement}
\label{sec:stages-7-8}

The ultimate purpose of \prometheus culminates in the stages of
validation and refinement, in which it seeks to empirically validate
that the selected domain heuristics are more effective than a control
group of heuristics. As explained in~\Cref{sec:prometheus},
\prometheus defines quantitative quality indicators for this
comparison, and to guide the process of iterative refinement if
necessary. Of course, it is always advisable to perform additional
validations, especially qualitative; \prometheus provides a solid
foundation for validating domain heuristics.

\parhead{Input} The selected domain heuristics, along with their
formal description. Additionally, the set of control heuristics to be
used. By default Nielsen’s heuristics~\cite{JackobNielsen1995} should
be used as a control group, unless there are more specific heuristics
and they are already validated.

\parhead{Output} After the experimental evaluation, the following
artifacts are obtained:

\begin{itemize}

\item Set of problems detected in the heuristic evaluation(s). The
  problems are classified into common problems, \problemasComunes,
  problems specific to the domain heuristics, \problemasDominioUniq,
  and problems specific to the control heuristics,
  \problemasControlUniq. Each problem is associated with a particular
  heuristic.

\item Each problem’s severity is specified, using a 5-level Likert
  scale.

\item For each problem in
  $\problemasComunes = \problemasComunes \cup \problemasControlUniq$,
  a specificity coefficient, with the same kind of scale as for
  severity.

\end{itemize}

In addition, with this information the following quality indicators
are calculated:

\begin{itemize}

\item Unique problem rate \tasaUnicidad.

\item Problem dispersion rate \tasaDispersion.

\item Problem severity rate \tasaSeveridad.

\item Problem specificity rate \tasaEspecificidad.

\end{itemize}

Regarding the indicators, those were described
in~\Cref{sec:prometheus}, and we only need to define the rate of
specificity. Let us recall that this indicator is defined as follows:

\[ \tasaEspecificidad = \especificidadDominioUniq / \especificidadControlUniq \]
 
\noindent
where \especificidadDominioUniq and \especificidadControlUniq are the
specificity averages for the problems found, respectively, by the
domain heuristics and the control heuristics.

The point we must explain is that the \especificidadDominioUniq and
\especificidadControlUniq calculations are different, although they
aim to quantify the same phenomenon: how specific the problems
encountered are. The problem is that when the control heuristics are
very general, for example Nielsen’s, it is difficult to decide when a
problem is domain-specific. For that reason, the problems in
\problemasControl need to be qualified according to specificity. That
is to say, \especificidadControlUniq corresponds to the simple
specificity average for each problem. Nevertheless, on the other hand,
the domain heuristics have been designed and selected based on their
specificity indices – in particular \fsi, which synthesizes this
characteristic on a heuristic level. Therefore,
\especificidadDominioUniq corresponds to a weighted sum of the number
of problems in each heuristic, multiplied by the \fsi associated with
it. In summary, we have that:

\begin{align*}
\especificidadDominioUniq &= \frac{\sum_{H_j} (|P \in H^D_j| * \fsi_j)}{|\problemasDominio|}\\
\especificidadControlUniq &= \frac{\sum_{i = 1}^n \varepsilon (P_i)}{|\problemasControl|}\\
\end{align*}

\noindent
for all domain heuristics $H_i$, where:

\begin{itemize}

\item $|P \in H_j|$ is the number of problems associated with said heuristic.

\item $|\problemasDominio|$ and $|\problemasComunes|$ correspond to
  the number of domain and control problems, respectively.

\item $\fsi_j$ is the final specificity index for said heuristic
  $H_j$.

\item $n$ is the total number of problems in \problemasControl.

\item $\varepsilon (P_i)$ is the individual specificity of a given
  problem $P_i$, and associated with the control group problems.

\end{itemize}

\section{Interview for the Initial Evaluation of \prometheus}
\label{sec:interv-init-eval}

In this section we present the interview conducted as part of the
initial validation of \prometheus, designed to collect information in
respect to the perception of \prometheus as a methodology to create
usability heuristics, based on the following aspects and whose
questions are summarized in \Cref{tab:entrevista}.

\begin{itemize}

\item Method of application
\item Interest or final product
\item Stages applied 
\item Easiest and/or most difficult stages
\item Usefulness  
\item Quantification of application time
\item Validation mechanisms of the generated products
\item Future plans for application of the process

\end{itemize}

\begin{table*}[!ht]
  \begin{scriptsize}
  \begin{center}
  \caption{Questions from the interview executed in the initial
    validation of \prometheus}
  \label{tab:entrevista}
  \begin{tabulary}{0.8\textwidth}{|L|L|}
    \hline
    Question
    & Purpose\\

    \hline
    ¿What is your subject of study?
    &  Determine the specific domain in which the researcher is working\\

    \hline
    What product or products do you expect to produce in your research?
    & Understand whether the researcher is developing new heuristics,
      refining existing ones, or making a detailed descripcion of heuristics.\\
    
    \hline
    Which stages of \prometheus do you think are applicable to your research?
    investigación?
    & Determine the potential applicability of stages in \prometheus
      to the particular project of the researcher\\
    
    \hline
    Considering all potentially applicable stages, which ones do you
    consider would be the easiest to apply?
    & Determine the ease of application of applicable stages of
      \prometheus\\
    
    \hline
    Considering all potentially applicable stages, which ones do you
    consider would be the hardest to apply?
    & Determine which applicable stages of \prometheus are considered as difficult to
      apply\\
      
    \hline
    Do you consider it necessary to add new stages or activities to
    the methodology?
    & Obtain experts' opinion about the proper quantity of stages or
      activities in \prometheus\\

    \hline
    Do you consider it necessary to remove or consolidate stages or
    activities in the methodology?
    & Obtain experts' opinion about the proper quantit of stages or
      activities in \prometheus\\
    
    \hline
    Do you understand in general terms how to apply the methodology?
    &  Identify the perception regarding the general clarity of the process\\

    \hline
    Do you understand how to apply each individual stage?
    &  Identify the perception regarding the clarity of the
      description of each stage of the process\\
    
    \hline
    Do you consider in general terms that the methodology is useful,
    as well as each of its stages?
    &  Identify the perception regarding the usefulness of the
      methodology\\

    \hline
    Do you consider that the quality indicators defined in \prometheus
    are relevant and points towards an effective quantitative validation?
    & Obtain experts' opinion regarding the pertinence of the quality indicators\\

    \hline
    Do you consider that the quality indicators defined in \prometheus
    are easy to compute?
    & Obtain experts' opinion regarding the ease of computation of the
      quality indicators\\

    \hline
  \end{tabulary} 
  \end{center}
  \end{scriptsize}
  
\end{table*}

\end{document}